\documentclass[12pt]{amsart}
\usepackage{mathrsfs}

\usepackage{amssymb,amsfonts,amsthm,amsmath}
\usepackage{epsfig}
\usepackage[utf8]{inputenc}
\usepackage{mathrsfs}
\usepackage{lscape}
\usepackage{amssymb,amsmath,graphicx,color,textcomp, amsthm,bbm,bbold, enumerate,booktabs}
\usepackage{chngcntr}
\usepackage{apptools}
\AtAppendix{\counterwithin{theorem}{section}}
\definecolor{Red}{cmyk}{0,1,1,0}

\definecolor{verde}{cmyk}{1,0,1,0}

\definecolor{loka}{cmyk}{.5,0,1,.5}
\definecolor{azul}{cmyk}{1,1,0,0}

\numberwithin{equation}{section}

\newcommand{\be}{\begin{equation}}
\newcommand{\ee}{\end{equation}}

\newtheorem{remark}{Remark}

\begin{document}
\title{Validation of a fractional model for erythrocyte sedimentation rate}
\author{J. Vanterler da C. Sousa$^1$}
\address{$^1$ Department of Applied Mathematics, Institute of Mathematics,
 Statistics and Scientific Computation, University of Campinas --
UNICAMP, rua S\'ergio Buarque de Holanda 651,
13083--859, Campinas SP, Brazil\newline
e-mail: {\itshape \texttt{vanterlermatematico@hotmail.com, capelas@ime.unicamp.br }}}
\author{Magun N. N. dos Santos}
\author{L. A. Magna}
\author{E. Capelas de Oliveira$^1$}

\address{Department of Clinical Pathology, School of Medical Sciences,\\
University of Campinas --UNICAMP, 13083-887, Campinas, SP, Brazil\newline
e-mail: {\itshape \texttt{magnun@fcm.unicamp.br}}}
\address{Department of Medical Genetics, School of Medical Sciences,\\
University of Campinas --UNICAMP, 13083-887, Campinas, SP, Brazil\newline
e-mail: {\itshape \texttt{amagna@uol.com.br}}}

\begin{abstract} We present the validation of a recent fractional mathematical model for erythrocyte sedimentation proposed by Sharma et al. \cite{GMR}. The model uses a Caputo fractional derivative to build a time fractional diffusion equation suitable to predict blood sedimentation rates. This validation was carried out by means of erythrocyte sedimentation tests in laboratory. Data on sedimentation rates (percentages) were analyzed and compared with the analytical solution of the time fractional diffusion equation. The behavior of the analytical solution related to each blood sample sedimentation data was described and analyzed. 

\end{abstract}
\maketitle

\section{Introduction} 

The study of differential equations with integer order has always proved
important and useful in mathematics, in particular, for the formulation of
mathematical models \cite{I2,I3,I4,I5,I6}. Several such models are based on
ordinary or partial differential equations. However, for systems with many
variables it is usually very difficult to build models that match reality.

Modeling such complex systems by means of a differential equation of
non-integer order presents some advantages as compared with mathematical models
that use only classical, integer order operators. Indeed, it has been realized
that the use of differential equations of fractional order to model certain
complex phenomena usually provides better descriptions of their behavior,
allowing us to obtain more accurate information about the underlying  physical
systems \cite{F1,F2,F3,F4,F5,F6}. Thus, fractional calculus became popular
because of its importance and relevance, specifically, due to its numerous
applications in several fields of science and particularly in biology and
medicine, in which the authors Sousa et al. \cite{JLEC}, using the Caputo
fractional derivative, introduced a fractional mathematical model that
describes the concentration of nutrients in blood, a parameter that affects
Erythrocyte Sedimentation Rate (ESR) and which contain, as a particular case, the result obtained in \cite{GMR}.

Due to this popularity of fractional calculus, there are numerous recent works
related to biology and medicine \cite{FM1,FM2,FM3,FM4,pahnehkol}. The
motivation for this work came from a mathematical model proposed by Sharma et
al. \cite{GMR}. First, Sousa et al.  \cite{JLEC}, using the Caputo fractional
derivative, proposed a fractional version of the model by Sharma et
al. \cite{GMR}, in order to generalize it and, possibly, make it 
more accurate. In order to obtain more accurate model information with respect
to reality, the model must be tested and verified, that is, by ESR test, look
at the data and compare with the analytical solution of the time fractional
diffusion equation.

The paper is organized as follows: In section 2, we present a brief explanation
about ESR, together with some tables showing analytical and clinical factors
that affect erythrocyte sedimentation, increasing and decreasing ESR. A brief
description of the erythrocyte sedimentation process and its three phases,
\textit{rouleaux} formation, precipitation and packaging closes the section. In
section 3, we present our fractional mathematical model and the corresponding
initial and boundary conditions. The analytical solution of the fractional
model is presented and the solution of the mathematical model proposed by
Sharma et al. \cite{GMR} is recovered.  In section 4 we present the experiments
conducted in order to obtain experimental values (percentages) of ESR for 8
blood samples from 8 different individuals, which would be compared with the
fractional mathematical model.  The results are presented in two complete
tables. In section 5, the main result of this paper, we validate the fractional
mathematical model using data from the erythrocyte sedimentation tests of the 8
samples. We present some graphs and discuss the results obtained.
Concluding remarks close the paper.


\section{Erythrocyte Sedimentation Rate}

ESR is a classic clinical test that measures how far erythrocytes settle into
the bottom of a test tube over a 1 hour period. The test was originally
described in 1897 by Biernacki in Poland \cite{EKU,EJK,EB,EB1}. It was
introduced by Fahraeus and Westergren at the beginning of the nineteenth
century \cite{AW,AW1,RF,RF1}.

ESR is in fact a very imprecise test due to the influence of analytical factors
such as temperature, table slope and stability, and clinical factors like
anemia, giant cells, diabetes, AIDS, smoking, drinking, weight and even height,
which can give rise to false-positive and false-negative results
\cite{NEL,JDA,SEB}.

Among clinical factors, the erythrocytes themselves and influence of plasma proteins, associated with inflammation are the ones that most influence ESR test results \cite{IPHJ,JCPA}. Nevertheless, the role of ESR in clinical decision making under non-characteristic conditions has been reestablished in different settings, including rheumatology, hematology and even orthopedics \cite{HP4,HP5,HP6,HP7,HP8,HP9,HP10}. 

The erythrocyte sedimentation test for red blood cells is performed within a
vertical 200 mm blood test tube. The following two tables present the clinical
factors that increase and decrease ESR. For the preparation of Table 1 we used
the papers
\cite{BV2,BV3,BV4,BV5,BV6,BV8,BV9,BV10,BV11,BV12,BV14,BV15,BV16,BV17}.

\begin{eqnarray*}
&&\text{ \ \ \ \ \ \ \ \ Table 1 - Factors influencing Erythrocyte Sedimentation.}  \\ 
&&%
\begin{tabular}{|c|c|} \hline
ESR increases & ESR decreases \\ \hline
Pregnancy & Polycythemia \\ \hline
Old age & Dysfibrinogenemia and afibrinogenemia \\ \hline
Anemia & Intravascular Diffuse Coagulation \\ \hline
$%
\begin{array}{c}
\text{High concentration of} \\ 
\text{non-fibrogenic proteins}
\end{array}%
$ & Congestive heart failure\\ \hline
Renal Insufficiency & Valproic Acid \\\hline 
Heparin & Cachexia \\ \hline
Rheumatoid Arthritis & Sickle Cell Disease \\ \hline
Tuberculosis & Hereditary Spherocytosis \\\hline 
Acute Infections & Hyperglycemia \\\hline 
Kidney Disease & Acanthocytosis \\ \hline
Macrocytosis & Microcytosis\\ \hline
Dextran & Spherocytosis\\ \hline
Diabetes Mellitus & Thalassemia \\ \hline
Gout & Cortisone \\ \hline
Multiple Myeloma & Quinine \\ \hline
Myocardial Infarction &  \\\hline 
Rheumatic Fever &  \\ \hline
Syphilis &  \\ \hline 
Temporal Arteritis  & \\ \hline
\end{tabular}
\end{eqnarray*}

On the other hand, it is important to have in mind that ESR values are
different for men, women, children and elderly people. Table 2 presents the
reference values of ESR for each type of individual.  For the preparation
of Table 2 we used references \cite{BO1,BO2,BO3,BO4,joi}.
\begin{eqnarray*}
&&\text{\ \ \ Table 2 - Normal values of ESR} \\ 
&&%
\begin{tabular}{|c|c|c|}\hline
Author & Patient & Values in mm \\ \hline
& Children & 4-7 \\ \hline
Miller & Men & 3-5 \\ \hline
& Women & 4-7 \\\hline 
&  &  \\ \hline
& Children & 0-10 \\ \hline
Borges & Man & 0-15 \\ \hline
& Women & 0-20 \\ \hline
\end{tabular}%
\end{eqnarray*}

Erythrocytes usually form aggregates, called \textit{rouleaux}, which resemble piles of coins. However, the formation of \textit{rouleaux} does not occur in the blood flow of a healthy human.

The erythrocyte sedimentation process can be divided in three phases: \textit{rouleaux} formation, precipitation and packaging. \textit{Rouleaux} formation is the most influential phase in determining the test result.
Normally, red blood cells have negative charges on their surfaces and repel each other, while many plasma proteins have positive charges and neutralize the surface charges of erythrocytes, promoting aggregation.
Thus, an increase in plasma proteins will be associated with higher ESR. Precipitation is the second phase of the erythrocyte sedimentation process; it occurs over a period of 40 min. Aggregates of erythrocytes fall under the
influence of gravity at a constant rate. Large aggregates fall faster than small aggregates and isolated cells. The falling aggregates induce an upward plasma current that delays sedimentation. Finally, the packaging stage takes place during 10 min. The sedimentation rate decreases and cells begin to pack at the bottom of the tube. In this way, the erythrocyte sedimentation process is terminated \cite{RO1,RO2}.


\section{Fractional mathematical model}

In this section we describe, as a review, the fractional mathematical model by means of the Caputo fractional derivative and the corresponding analytical solution proposed by Sharma et al. \cite{JLEC}.

The concentration of nutrients in blood is a function $C(x,t)$ twice continuously differentiable that satisfies the following non-homogeneous time-fractional partial differential equation (PDE),
\begin{equation}  \label{S1}
D_{L}\mathcal{D}_{x}^{2}C\left( x,t\right) - \mathcal{D}_{t}^{\mu }C\left(
x,t\right) =\phi \left( x,t\right) ,
\end{equation}
where $\mathcal{D}^{\mu}_t \equiv \displaystyle\frac{\partial^{\mu }}{\partial t^{\mu}}$ is the Caputo fractional derivative, with $0<\mu\leq 1$, $D_{L}$ is a positive constant and $\phi(x,t)$ is a function twice continuously differentiable describing the nutrient transfer rate and which satisfies the PDE
\begin{equation}  \label{S2}
D\mathcal{D}_{x}^{2}\phi \left( x,t\right) -k\phi \left( x,t\right) -
\mathcal{D}_{t}\phi \left( x,t\right) =0,
\end{equation}
with $D$ and $k$ both positive constants.

The initial and boundary conditions imposed here are given by
\begin{equation*}
\left\{ 
\begin{array}{ll}
\phi (x,0)=\exp \left( -\sqrt{\frac{k-a}{D}}x\right) , & k\geq a,D>0 , \\ 
\phi (0,t)=\exp \left( -at\right) , & t>0 , \\ 
\phi (\infty ,t)=0, & t>0.
\end{array}%
\right.
\end{equation*}

The solutions of the Eq.(\ref{S2}) can be written as
\begin{equation*}
\phi \left( x,t\right) =\exp \left( -\left( at+bx\right) \right),
\end{equation*}
where $b^{2}=\displaystyle\frac{\left( k-a\right) }{D}>0$ and $a$ is a constant to be adequately
chosen from a known value of $\phi \left( x,t\right)$.

Furthermore, we must impose the initial and boundary conditions for Eq.(\ref{S1}): 
\begin{equation}
\left\{ 
\begin{array}{ll}
C(x,0)=0, & x\geq 0 \\ 
C(0,t)=1, & t>0 \\ 
C(\infty ,t)=0, & t>0,
\end{array}%
\right.  \label{S4}
\end{equation}
with $C(x,t)\in C^{2}[0,b]$.

Thus, from these considerations, it follows that the time-fractional mathematical model to be addressed is composed of a non-homogeneous fractional PDE
\begin{equation}  \label{S3}
D_{L}\mathcal{D}_{x}^{2}C\left( x,t\right) - \mathcal{D}_{t}^{\mu }C\left(
x,t\right) =\exp \left( -\left( at+bx\right) \right) , \quad a, b \in \mathbb{R}, 
\end{equation}
with initial and boundary conditions given by Eq.(\ref{S4}) and $0 <\mu \leq 1$

We solve this problem, employing the methodology of Laplace transform to convert the non-homogeneous fractional PDE into a non-homogeneous linear ordinary differential equation.

Thus, the solution associated with our problem, i.e., a solution of Eq.(\ref{S3}) satisfying the conditions given by Eq.(\ref{S4}), is \cite{JLEC}
\begin{eqnarray}  \label{S43}
C\left( x,t\right) &=&t^{\mu }\underset{m=0}{\overset{\infty }{\sum }}\frac{%
\left( -\alpha xt^{-\mu /2}\right) ^{m}}{m!}\underset{k=0}{\overset{\infty }{%
\sum }}\left( -at\right) ^{k}\mathbb{E}_{\mu ,\mu +k+1-\mu m/2}\left( \beta
^{2}t^{\mu }\right) +  \\
&&+\mathbb{W}\left( -\mu /2,1;-\frac{\alpha x}{t^{\mu /2}}\right) -\exp
\left( -bx\right) t^{\mu }\underset{k=0}{\overset{\infty }{\sum }}\left(
-at\right) ^{k}\mathbb{E}_{\mu ,\mu +k+1}\left( \beta ^{2}t^{\mu }\right) \notag ,
\end{eqnarray}
where $\mathbb{E_{\mu,\gamma}}(\cdot)$ is the two-parameter Mittag-Leffler function, $\mathbb{W}(a,b;\cdot)$ is the Wright function \cite{GKAM}, and the parameters are given by $\alpha ^{2}=1/D_{L}$, $\beta ^{2}=b^{2}D_{L}$ and $0<\mu \leq 1$. The solution given by Eq.(\ref{S43}) is $AC^{n}[0,b]$ and class $C^{2}$, then substituting it in Eq.(\ref{S3}) and evaluating the calculation we verify that it satisfies the initial value problem \cite{JVC}.

Note that, the solution of the fractional PDE in the limit $\mu \rightarrow 1$, recover the result by Sharma et al. \cite{GMR}.

\section{Experimental} 
\subsection{Materials and methods}

The ESR tests were conducted at the Hematology Laboratory of the Clinical Hospital, State University of Campinas, state of São Paulo, southeast Brazil, during the period of August 14-15, 2017, all of them between 2:30 pm and 4:00 pm. For the erythrocyte sedimentation test, the Westergren \cite{AW,AW1,RF,RF1} method was adopted as the reference method for ESR measurement by the international council of ICSH \cite{ICH,ICH1}. On this date, 8 samples were collected, 4 males and 4 females. We used a test tube of length 200 mm, but for evaluation purposes, we consider up to 120 mm height with an ambient temperature of 22º C and the angle between the tube and the table is 90º.

Patient blood samples were placed in test tubes containing no anticoagulant in the 8 samples. The height of the erythrocyte column in the test tube, in millimeters, was recorded at a 5 min interval for 1 hour, and was subsequently converted to sedimentation rate (percentage) according to the decreased proportion in height. The following tables indicate the values of ESR and clinical factors of the tests carried out on the 8 samples collected. Table 3, below, describes the sedimentation rate of each individual as time progresses. A detailed description of some information from the individuals in whom erythrocyte sedimentation tests were performed are shown in Table 4. An important observation related to Table 4 is the fact that besides presenting the ESR values (manual sedimentation), we also present the sedimentation performed in the automated system.

The reason for presenting these two versions, made manually and by device, is that the version via apparatus changes the properties of the blood and consequently the sedimentation. In this sense, it is necessary and important to present the difference between the two corresponding rates.

\begin{landscape}
\begin{eqnarray*}
&&\text{ \ \ \ \ \ \ \ \ \ \ \ \ \ \ \ \ \ \ \ \ \ \ \ \ \ \ \ \ \ \ \ \ \ \
\ \ \ \ \ \ \ Table 3 - Sedimentation values in the time intervals.} \\
&&%
\begin{tabular}{|c|c|c|c|c|c|c|c|c|c|c|c|c|c|c|} \hline
Time/Pipe & 0 & 5 & 10 & 15 & 20 & 25 & 30 & 35 & 40 & 45 & 50 & 55 & 60 & 
\\ \hline
Pipe 1 - F & 0 & 0.83 & 0.83 & 0.83 & 1.66 & 3.32 & 6.64 & 9.96 & 13.28 & 
16.60 & 19.09 & 24.07 & 25.73 &  \\ \hline
Pipe 2 - F & 0 & 0.83 & 0.83 & 0.83 & 1.66 & 2.49 & 3.32 & 4.15 & 5.81 & 7.47
& 9.96 & 11.62 & 14.11 &  \\ \hline
Pipe 3 - F & 0 & 0.83 & 0.83 & 0.83 & 2.49 & 4.98 & 9.96 & 14.11 & 20.75 & 
29.05 & 34.86 & 40.67 & 45.65 & E \\ \hline
Pipe 4 - F & 0 & 1.66 & 1.66 & 2.49 & 3.32 & 3.32 & 8.30 & 19.09 & 30.71 & 
41.50 & 51.46 & 59.76 & 65.57 & S \\ \hline
Pipe 1 - M & 0 & 0 & 0.83 & 0.83 & 1.66 & 1.66 & 2.49 & 2.49 & 4.15 & 5.81 & 
7.47 & 9.96 & 12.45 & R \\ \hline
Pipe 2 - M & 0 & 0 & 0.83 & 0.83 & 0.83 & 0.83 & 1.66 & 1.66 & 2.49 & 3.32 & 
4.98 & 5.91 & 6.64 &  \\ \hline
Pipe 3 - M & 0 & 0.83 & 1.66 & 2.49 & 4.15 & 7.47 & 12.45 & 18.26 & 23.24 & 
29.88 & 35.69 & 40.67 & 45.65 &  \\ \hline
Pipe 4 - M & 0 & 0 & 0 & 0 & 0.83 & 0.83 & 0.83 & 0.83 & 0.83 & 1.66 & 1.66
& 1.66 & 2.49 & \\ \hline
\end{tabular}
\end{eqnarray*}
\end{landscape}

\begin{landscape}
\begin{eqnarray*}
&&\text{ \ \ \ \ \ \ \ \ \ \ \ \ \ \ \ \ \ \ \ \ \ \ \ \ \ \ \ \ \ \ \ \ \ \ \ \ \ \ \ \ \ \ \ \ \ \ \ \ Table 4 - Patient information.} \\
&&%
\begin{tabular}{|c|c|c|c|c|c|c|c|c|}\hline
Label & Sex & Height & Weight & BMI & Man. Sed. & Apa. Sed. & Disorder & Diff.
\\ \hline
1 & F & 1.60 m & 72.00 kg & 28.1 & 24.73 \% & 27.00 \% & Bursitis Trochanterica
 & 9.18 \%  \\ \hline
2 & F & 1.58 m & 65.00kg & 26.0 & 14.11 \% & 33.00 \% & Spondylarthritis
 & 
133.88 \% \\ \hline
3 & F & 1.52 m & 42.70 kg & 18.5 & 45.65 \% & 50.00 \% & Control
 & 9.53 \%
\\ \hline
4 & F & 1.58 m & 81.00 kg & 32.4 & 65.57 \% & 68.00 \% & Polyarthralgia
 & 
3.71 \% \\ \hline
5 & M & 1.67 m & 78.50 kg & 28.1 & 12.45 \% & 19.00 \% & Rheumatoid arthritis & 52.61 \% \\ \hline
6 & M & 1.90 m & 105.0 kg & 29.1 & 6.64 \% & 3.00 \% & Lupus & -54.82 \% \\ \hline
7 & M & 1.69 m & 74.00 kg & 25.9 & 45.65 \%  & 39.00 \% & Retinopathy
 & 
-14.57 \% \\ \hline
8 & M & 1.87 m & 73.00 kg & 20.9 & 2.49 \% & 3.00 \% & $
\begin{array}{c}
\text{Vasculitis} \\\hline 
\text{Lupus Erythematosus} \\ \hline
\text{Disseminated}
\end{array}%
$ & 20.48 \% \\ \hline
\end{tabular}
\end{eqnarray*}
\end{landscape}

\begin{remark}
\begin{itemize}\
\item For individuals 1, 2 and 3: unlikely to influence ESR;
\item For individuals 4 and 7: may influence ESR;
\item For individuals 5, 6 and 8: do influence ESR.
\end{itemize}
\end{remark}

\section{Results and Discussion} 

In this section, we will present the graphs of the results obtained in the ESR
realization and make some analysis and comparisons with the solution of the
fractional diffusion equation Eq.(\ref{S1}). Discussions involving the
sedimentation rate of erythrocytes, the data collected and the solution of the
fractional diffusion equation, conclude the section. 

We start with the analysis of Graphs 1-4, which represent the sedimentation rate of four males.  

\begin{figure}[ht!]
\caption{ESR person 1: Male.}
\label{fig:eryt1}\centering 
\includegraphics[width=12cm]{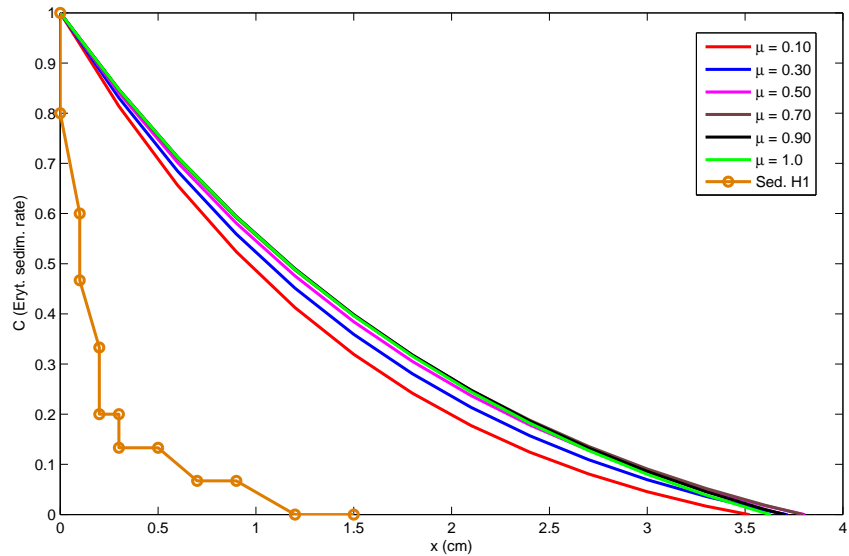}
\end{figure}

\begin{figure}[ht!]
\caption{ESR person 2: Male.}
\label{fig:eryt1}\centering 
\includegraphics[width=12.5cm]{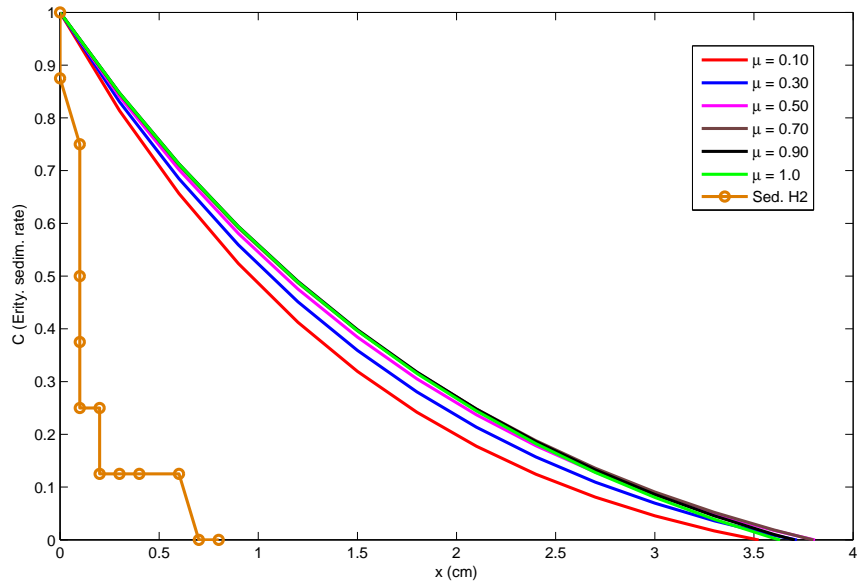}
\end{figure}
\newpage

\begin{figure}[ht!]
\caption{ESR person 3: Male.}
\label{fig:eryt1}\centering 
\includegraphics[width=12cm]{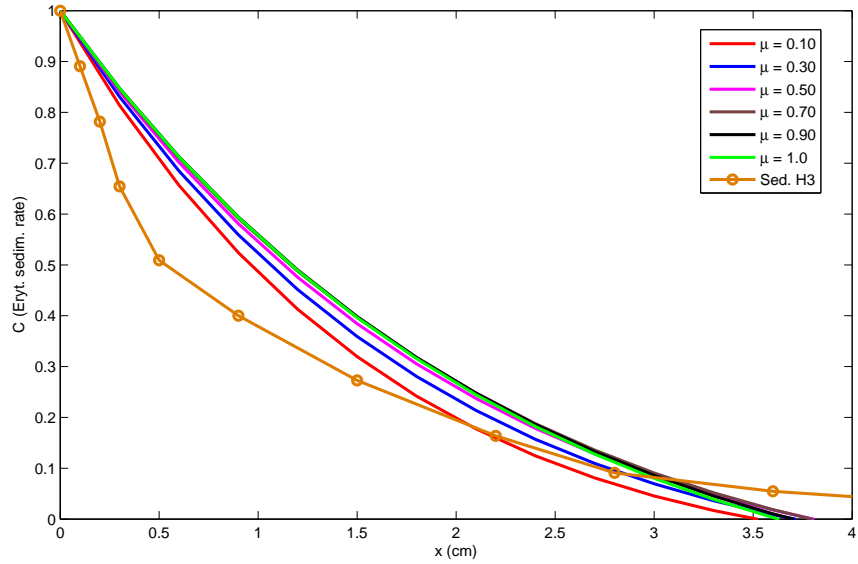}
\end{figure}

  \begin{figure}[ht!]
  \caption{ESR person 4: Male.}
  \label{fig:eryt1}\centering 
  \includegraphics[width=14cm]{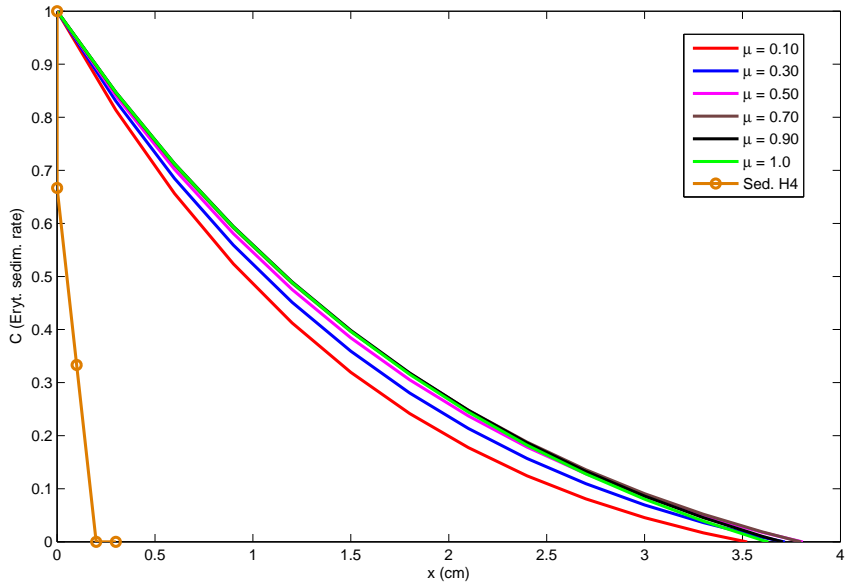}
  \end{figure}
  \newpage
  
With the purpose to look at the data of individuals, and plot their respective graphs, for a particular choice of the parameter $0<\mu\leq 1$, we tried to approach the analytic solution of the fractional diffusion equation and obtain more precise information on erythrocyte sedimentation behavior in that interval. Then, Graphs 1-4, allow to carry out an analysis of the sedimentation of erythrocytes. The graphs were plotted using the software MALTAB 6.10.

In order to look at the data and validate the fractional mathematical model, recently introduced by Sousa et al. \cite{JLEC}, the graphs present an expected behavior, in the sense that each individual has a certain clinical problem. The erythrocyte sedimentation behavior via a graphical analysis varies from individual to individual. Note that from the graph of male 1, the erythrocyte sedimentation behavior presents a characteristic similar to the solution of the fractional diffusion equation, that is, as it moves away from the border ($ x \neq 0 $), the concentration of nutrients decreases. On the other hand, the graphs of individuals 2 and 4, the characteristic of the sedimentation curve does not present a good behavior, this is due to the characteristics of each individual, from clinical factors  among them, the age. However, if we perform other experiments, we could obtain better results and consequently the behavior of the ESR graph is somehow best viewed by the analytical solution given by
Eq.(\ref{S43}). On the other hand, it is possible to note that the behavior of the analytical solution of Eq.(\ref{S43}) better approach the erythrocyte sedimentation of individual. In fact, it is possible to choose $n$ individuals and to study the graph that describes the ESR and thus to make the model more accurate.

The next Graphs 5-8, refer to the behavior of erythrocytes sedimentation for the other 4 individuals, in this case females.

\begin{figure}[ht!]
\caption{ESR person 1: Female.}
\label{fig:eryt1}\centering 
\includegraphics[width=13.5cm]{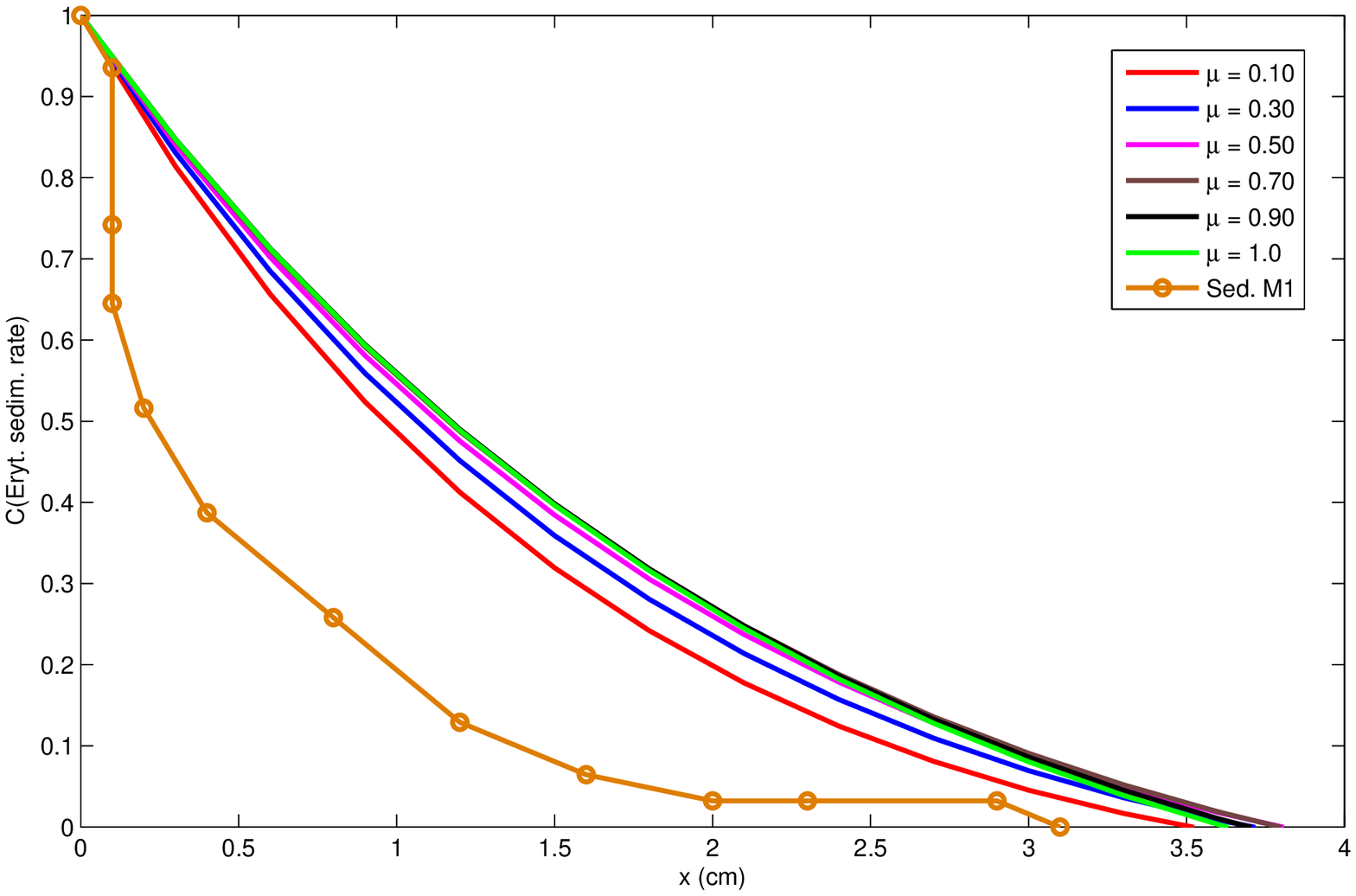}
\end{figure}

\begin{figure}[ht!]
\caption{ESR person 2: Female}
\label{fig:eryt1}\centering 
\includegraphics[width=12cm]{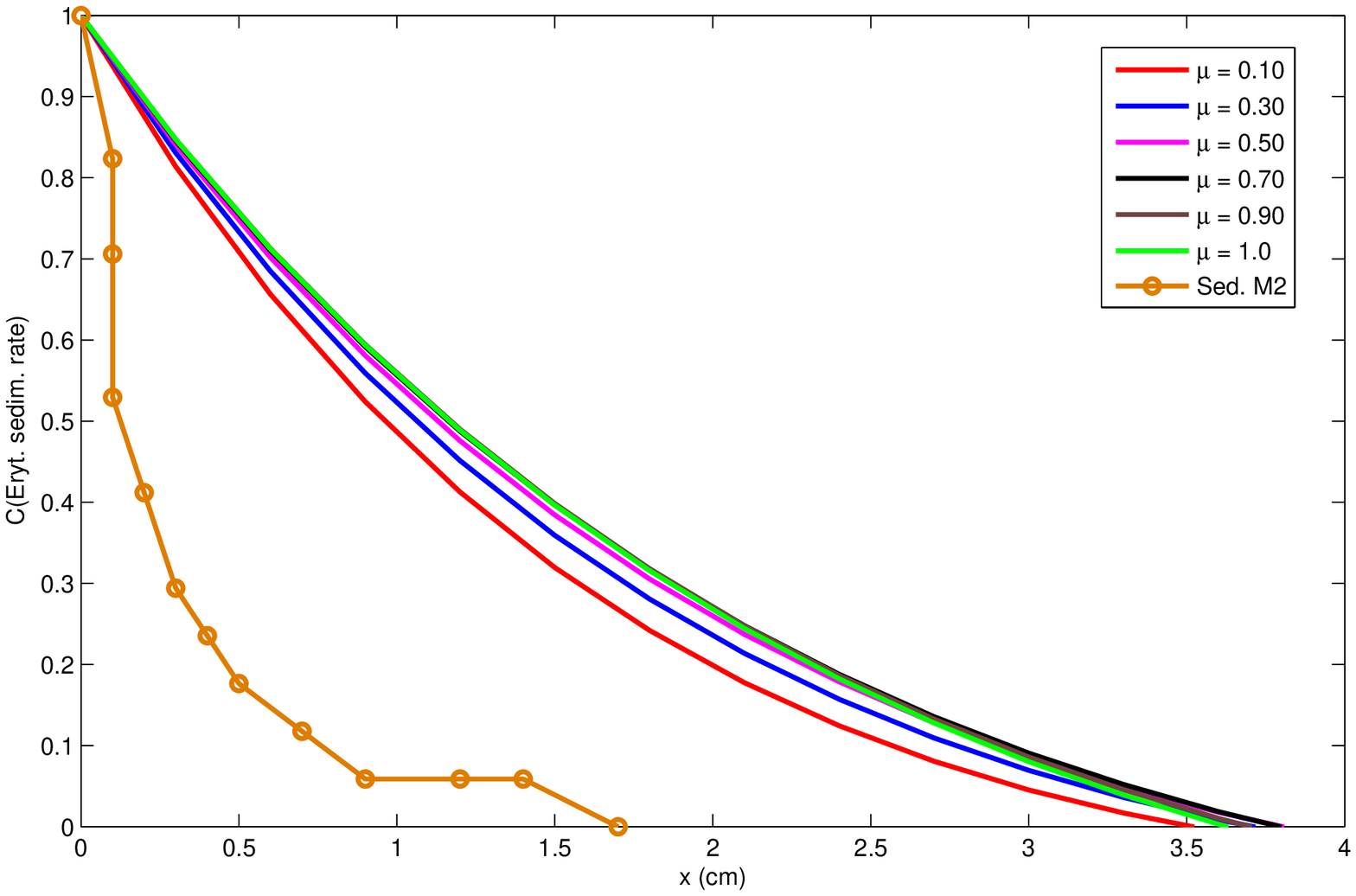}
\end{figure}
\newpage
\begin{figure}[ht!]
\caption{ESR person 3: Female}
\label{fig:eryt1}\centering 
\includegraphics[width=14.5cm]{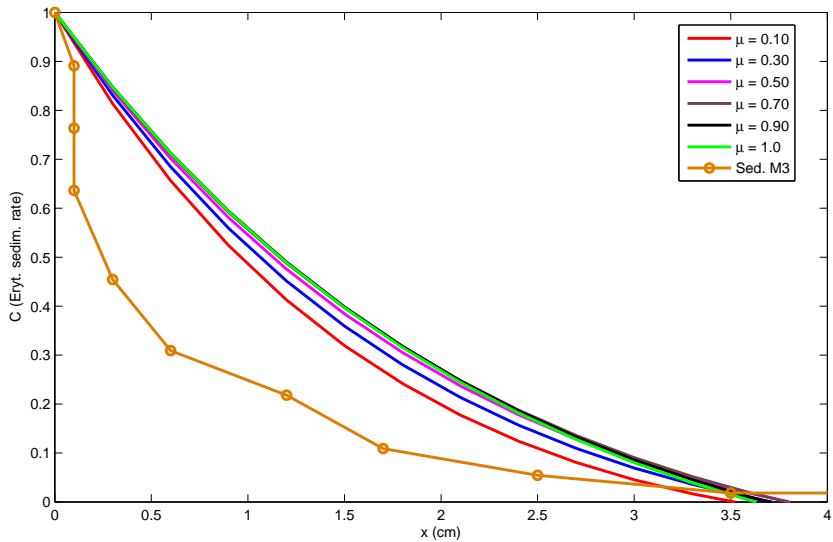}
\end{figure}

\begin{figure}[ht!]
\caption{ESR person 4: Female}
\label{fig:eryt1}\centering 
\includegraphics[width=13cm]{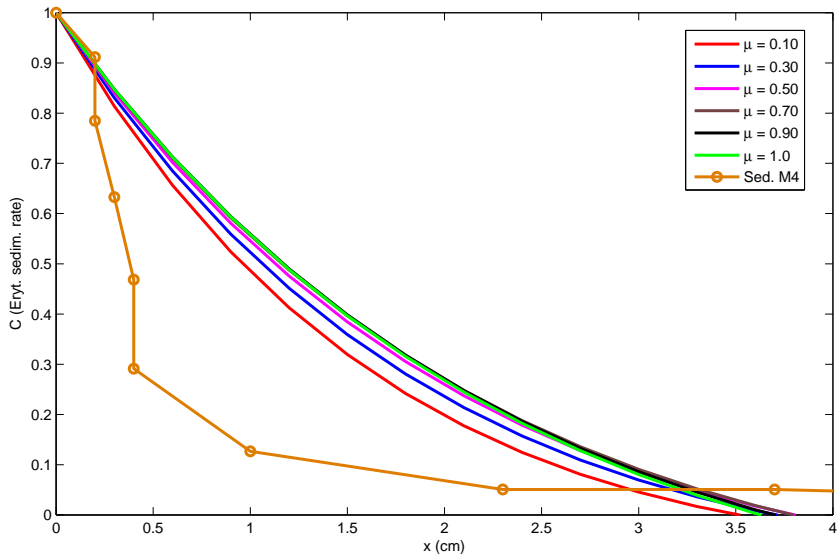}
\end{figure}

At first, it can be noted that the behavior of ESR of four female individuals presented in the four graphs above is in fact better and the solution of the fractional diffusion equation, the data tape better. However, in the same way that we can perform $n$ tests of female individuals and find others in which the solution Eq.(\ref{S43}) tape the data more efficiently, making the fractional mathematical model able to offer information closer to reality in regards to erythrocyte sedimentation. 

In addition, it is worth noting that, as observed in the Graphs 1-8 the fractional mathematical model proposed by Sousa et al. \cite{JLEC} via a Caputo fractional derivative, allows a variation in the order of the
derivative, and consequently in the analytical solution of the Eq.(\ref{S1}), making it possible to look better at the data. On the other hand, the mathematical model as proposed by Sharma et al. \cite{GMR} does not
allow this freedom to look at the data in the same way as our model. Thus, our fractional mathematical model allows a better reading of the data, that is, the concentration of nutrients in the blood, in relation to the  model proposed by Sharma et al. \cite{GMR}.


\section{Concluding remarks}

After a brief introduction to a fractional mathematical model used to describe the blood concentration of nutrients that influence the sedimentation rate of blood cells, we present the results of laboratory ESR tests with 8 samples, 4 males and 4 females, used to validate that mathematical model. For this sake, the results of sedimentation tests were compared, through graphical analysis, with the analytical solution of the mathematical model.  Only 8 samples were enough to carry out this comparison and discuss the model. However, more tests, providing new data, would surely allow a better comprehension of the fractional mathematical model, making it more appropriate to describe reality.

A remarkable advantage of the fractional mathematical model is the freedom  to choose the order of the derivative in it and, consequently, the analytical solution of the problem. 

With the freedom given to parameter $ 0 <\alpha \leq 1 $, we were able to analyze the concentration of nutrients on certain intervals and regions in which the mathematical model proposed by Sharma et al. \cite{GMR} cannot be
used. Our model can thus provide information on the blood concentration of nutrients closer to reality.

In a future work, we intend to consider another type of fractional derivative \cite{ZE1,ZE2,mae,mae1}, namely, a general problem of fractional space diffusion. After that we shall carry out the same sedimentation tests and shall use their results to analyze the capabilities of that new model for describing real living systems.

\section*{Acknowledgment}
We would like to thank the Dr. J. Emlio Maiorino for several suggestions and comments that helped improve the paper.

\bibliography{ref}
\bibliographystyle{plain}

\end{document}